# Mesoscopic Stacking Reconfigurations in Stacked van der Waals Film

*Yoon Seong Heo, Tae Wan Kim, Wooseok Lee. Jungseok Choi, Soyeon Park, Dong-Il Yeom, and Jae-Ung Lee\**

Y. S. Heo, T. W. Kim, W. Lee, J. Choi, S. Park, D.-I. Yeom, J.-U. Lee
Department of Physics and Department of Energy Systems Research, Ajou University,
Suwon 16499, Korea
E-mail: jaeunglee@ajou.ac.kr

Mesoscopic-scale stacking reconfigurations are investigated when van der Waals films are stacked. We have developed a method to visualize complicated stacking structures and mechanical distortions simultaneously in stacked atom-thick films using Raman spectroscopy. In the rigid limit, we found that the distortions originate from the transfer process, which can be understood through thin film mechanics with a large elastic property mismatch. In contrast, with atomic corrugations, the in-plane strain fields are more closely correlated with the stacking configuration, highlighting the impact of atomic reconstructions on the mesoscopic scale. We discovered that the grain boundaries don`t have a significant effect while the cracks are causing inhomogeneous strain in stacked polycrystalline films. This result contributes to understanding the local variation of emerging properties from moiré structures and advancing the reliability of stacked vdW material fabrication.

## 1. Introduction

Van der Waals (vdW) materials provide a unique platform for studying physical properties at atomically thin limits. Few-atom-thick crystals exhibit contrasting behaviors compared to their bulk counterparts, which can be further manipulated by artificial stacking to form heterostructures.[1–4] When two monolayer crystals are stacked, moiré superlattices emerge, which modulate their electrical and optical properties.[5–8] The moiré length depends on their relative twist angles, which can be used as a knob to tailor their properties even with the same material combinations.[3,9] The polymer-assisted transfer method is widely used to fabricate stacked vdW materials. Due to the elastic property mismatch, mechanical distortions are easily introduced during the fabrication of stacked structures. This leads to local structural variation such as rotated moiré domains or residual strains.[10] Furthermore, the competition between interlayer interaction and strain leads to interesting structural reconstructions that significantly affect their properties.[11–14] Such atomic reconstructions even occur on a mesoscopic scale, exhibiting continuously varying moiré domains with complicated strain fields.[15]

In this regard, identifying structural variations of stacked vdW films on the mesoscopic scale is important. Researchers have developed several methods to identify structures in stacked vdW materials. The most reliable method is based on transmission electron microscopy (TEM).[11] Although TEM can observe various structural changes with atomic resolution, it requires special sample preparation and is challenging to apply for large-area characterization. To overcome these limitations, scanning electron microscopy (SEM) and scanning probe microscopy (SPM) based techniques have been developed.[16,17] These techniques have advantages for studying heterostructures with relatively long moiré periodicity with nanoscale resolution. However, they can only be applied to a limited range of twist angles, and investigating detailed information on the strain field is difficult. In contrast, Raman spectroscopy has advantages as a non-destructive probe for studying the structural configurations and interlayer interactions of the vdW system. Several reports studied the Raman signatures in stacked-bilayer systems which are summarized in Table S1 (Supporting Information).[18–23] Twist-angle-dependent Raman studies have provided information on angle-dependent interlayer interactions, the superlattice effect of the moiré structures, and delicate changes in the strain field due to atomic rearrangements. This rich information from a single measurement tool opens the possibility of using Raman spectroscopy as a versatile characterization tool for stacked vdW materials.

In this paper, we studied mesoscopic stacking reconfiguration in stacked vdW systems using optical spectroscopy. We identified the origins of mechanical distortions in the fabricated samples by comparing stacking structures and in-plane strain fields. As a model system, we investigated the hexagonal transition metal dichalcogenides (TMDs), such as $WS_2$ and $MoS_2$, which are representatives of atomically thin semiconductors. The $WS_2$ (and $MoS_2$) films were grown by metal-organic chemical vapor deposition (MOCVD) and then stacked using a modified dry stamping technique.[24,25] We categorized the possible stacking configurations into four different regimes (highly relaxed, transition 1, transition 2, and rigid regimes) and correlated them with characteristic signatures of the phonons to visualize stacking configurations. Along with the visualization of the stacking orders, the strain field can be analyzed simultaneously from the in-plane Raman modes. We found signatures of delamination at the cracks in the films that cause rotated moiré domains. Moreover, the shear-lag effect and atomic reconstructions produce different built-in strains inside the samples, which need to be taken into account when studying stacked vdW materials. Our technique is advantageous for the non-destructive identification of local structural variations of vdW films on the mesoscopic scale, which is crucial for utilizing the unique properties of the stacked vdW materials.

## 2. Results and discussion

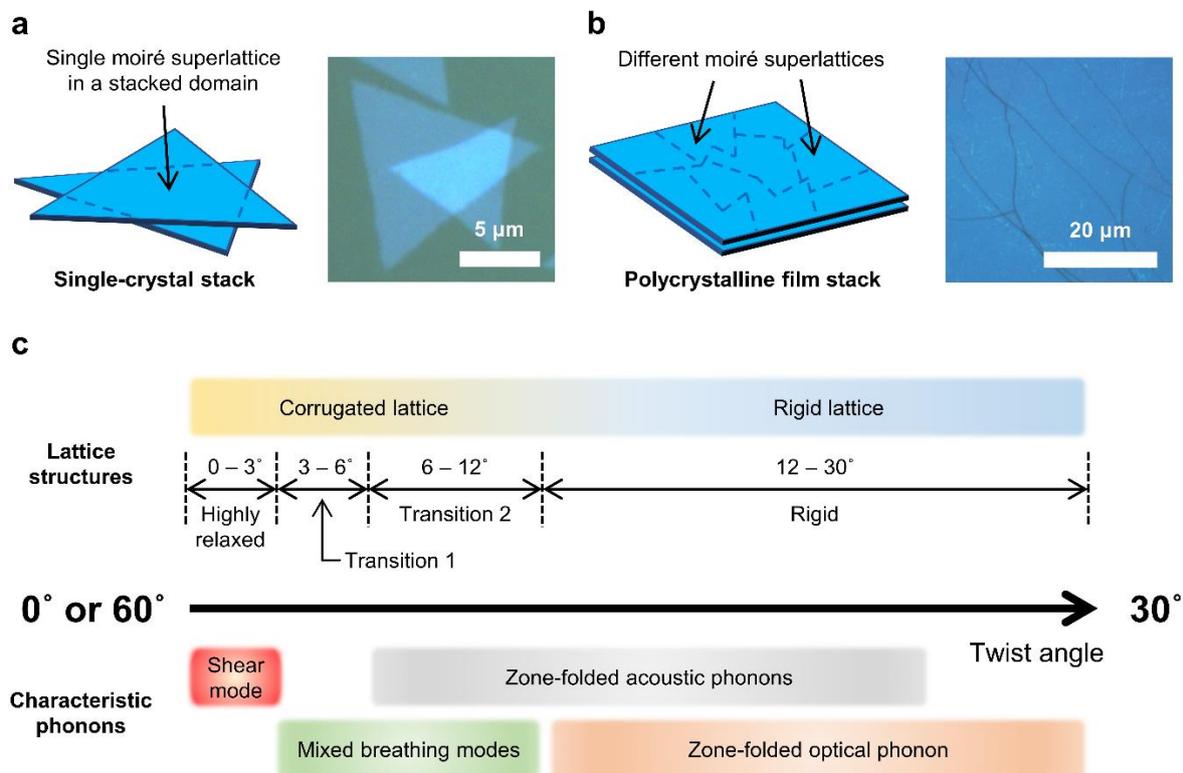

**Figure 1.** (a) Schematics and optical microscope (OM) images of stacked single-crystalline and (b) polycrystalline films. (c) Lattice structures and characteristic phonons as a function of the twist angles.

We fabricated stacked vdW films from either single-crystalline or polycrystalline films as shown in Figure 1(a,b). The MOCVD-grown single-crystal $WS_2$ has a triangular shape due to crystal symmetry and growth kinetics.[26] These single-crystalline film stacks can be used as a reference, as the relative angle can be identified by comparing optical microscope images and polarized second harmonic generation (SHG) measurements. (See Supporting Information, Note S1 and Figure S1) The typical error for the estimated twist angle is about 0.8°. Monolayer films synthesized on typical amorphous insulating substrates exhibit polycrystallinity. Nucleation occurs randomly on the surface unless it is controlled and crystals with different orientations are connected by grain boundaries.[27] When two polycrystalline films are stacked, the resulting moiré superlattices have random angle distributions with domain boundaries of moiré superlattices. Figure 1(c) shows the lattice structures when two monolayers are stacked. Depending on the twist angle, the moiré structure emerges due to the overlap of the two periodic structures. The moiré period depends on the twist angle, with a minimum value at 30°, increasing as it approaches 0° or 60°.[19] With a small moiré period, there are no significant distortions in the constituent layers, which corresponds to the rigid layer approximation. However, with a long moiré period, the atoms modulate their positions compared to the rigid approximation. This leads to an anomalous variation of the strain field in the constituent layer and atomic corrugations in the out-of-plane directions.[11,12]

Considering such differences as a function of the twist angle, we categorized the possible stacking configurations into four regimes based on the characteristic phonon signatures as shown in Figure 1(c). In hexagonal TMDs, there are two naturally occurring stacking orders called H- and R-type structures.[28] The R-type stacking has two layers aligned without twisting, corresponding to 0°. The H-type, which is the most stable stacking order for bilayers, has a relative angle corresponding to 60°. With small twist angles (0° to 3° and 57° to 60°), local high-symmetry domains (corresponding to R- or H-type stackings) are well-defined by lattice reconstruction. When the size of the high-symmetry domains is sufficient, the shear mode appears due to phonon renormalization and this is the main feature of a highly relaxed regime.[20,29] The transition regime from a corrugated to a rigid lattice is divided into two parts. In transition regime 1 (3° to 6° and 54° to 57°), the two Raman active breathing modes are present due to phonon-phonon coupling.[20,22] In transition regime 2 (6° to 12° and 48° to 54°), there is only one breathing mode that is redshifted compared to the rigid case due to out-of-plane corrugation.[23,29] With a small moiré period (12° to 48°), the system can be approximated with the rigid lattice. In this regime, the phonon structures can be described by the zone folding effect from its constituent layer.[19] The size of the Brillouin zone depends on the moiré period, so zone-folded acoustic and optical phonons shift as a function of the twist angle. Note that the zone-folded acoustic phonons start to emerge from transition regime 2 without any signatures of zone-folded optical phonons.

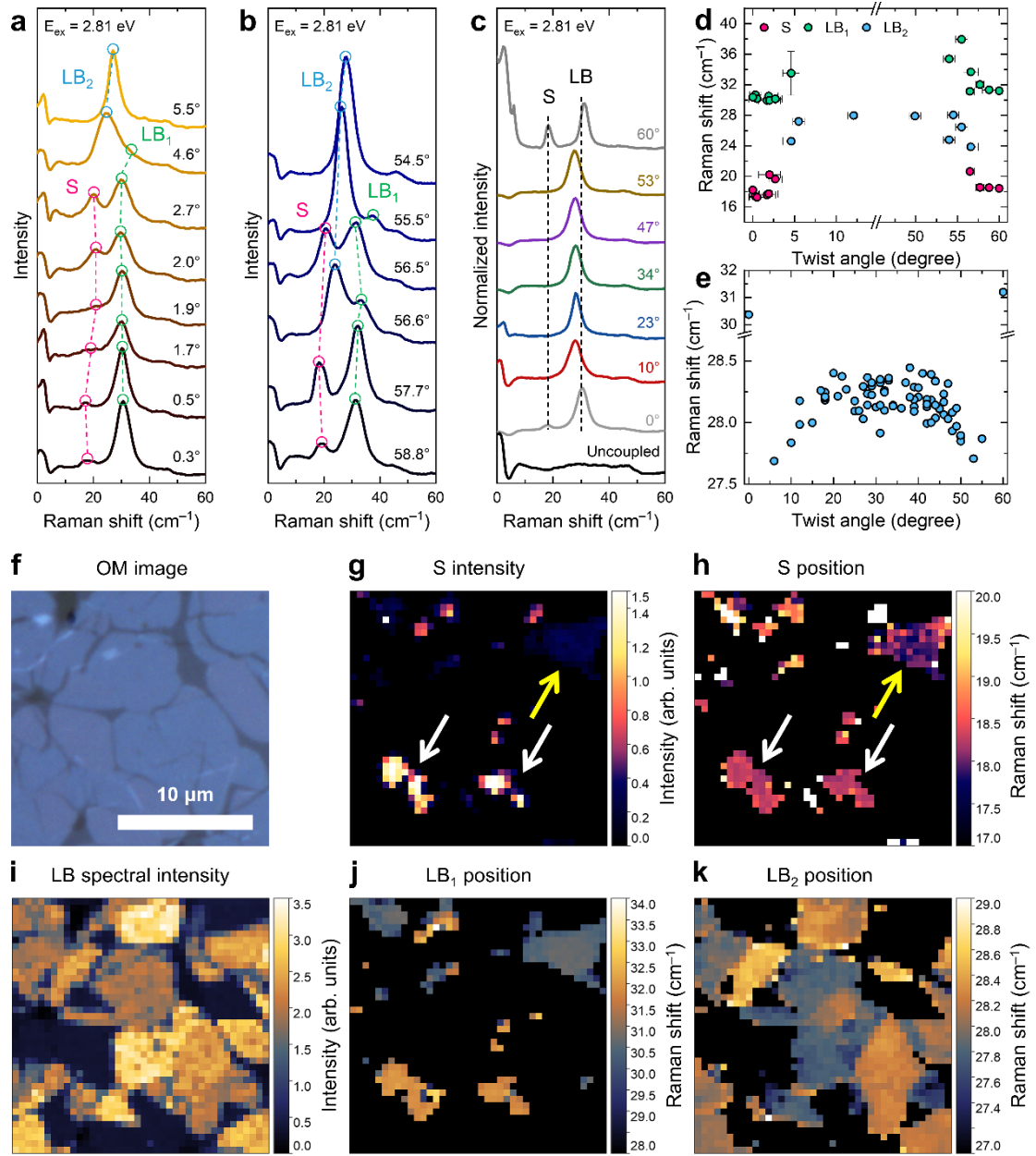

**Figure 2.** Interlayer Raman modes with twist angles ranging from (a) 0.3° to 5.5°, (b) 54.5° to 58.8°, and (c) other twist angles. The typical error for the estimated twist angle is about 0.8°. The Raman peak positions of the (d) corrugated (e) rigid structures. (f) The OM image of stacked polycrystalline samples. The Raman images of (g) S mode intensity, (h) S mode position, (i) the LB spectral intensity from 10 to 50 cm$^{-1}$, and the peak position of (j) LB$_1$ and (k) LB$_2$, respectively.

Low-frequency (<50 cm$^{-1}$) Raman spectroscopy is an unambiguous probe of interlayer interaction in layered materials. The shear (S) and layer breathing (LB) modes arise from vibrations of layers along in-plane and out-of-plane directions, respectively.[30] Since there are no intralayer atomic movements, these modes depend sensitively only on interlayer interactions. The interlayer Raman modes have been used to examine the interlayer spring constants, identify the number of layers and stacking orders, and study the shear deformation of the layered crystals.[30–32] Figure 2 shows the low-frequency Raman spectra of bilayer $WS_2$ with corrugated [Figures 2(a) and (b)] and rigid structures [Figure 2(c)]. We measured approximately 120 samples with different twist angles, as summarized in Figures 2(d) and (e). Even though the two layers are stacked, there are cases with no interlayer Raman modes, corresponding to the samples without sufficient interlayer coupling due to surface disorders or mechanical deformations. We performed Raman mapping on stacked polycrystalline films [Figure 2(f)] to investigate local variations as shown in Figure 2(g–k). Figure 2(i) shows the coupled and uncoupled areas, which correspond to the LB spectral intensity from 10 to 50 $cm^{-1}$. The presence of interlayer Raman modes is an excellent indicator to identify interlayer interactions.

With all twist angles, at least one LB mode is observable. For the twist angle of 0° and 60° (R- and H-type stacking, respectively), both the S and LB modes appear in the low-frequency Raman spectrum. The frequency of the S modes is similar for 0° (~18.2 $cm^{-1}$) and 60° (~18.4 $cm^{-1}$), while the LB mode is redshifted by 0.8 $cm^{-1}$ for the 0° (~30.4 $cm^{-1}$) compared to the 60° (~31.2 $cm^{-1}$). This reflects the difference in the interlayer distance between the two stable stacking orders.[29] For samples in a highly relaxed regime with a small twist angle, both the S and LB modes are observed. If we assume the moiré structure with rigid layers, the S modes should not appear at this twist angle. However, due to lattice reconstructions, local domains with high symmetry stackings (R- and H- types) are formed, which renormalize the energy of the shear phonons.[20] In addition, the detailed atomic structures of small-angle twisted bilayers TMDs close to 0° and 60° are different.[33] For the angle close to 0°, triangular R-type domains with inversions, which share identical lowest energy configurations, are formed along with partial dislocation boundaries. For the angles close to 60°, Kagome-like hexagonal patterns have been observed. Unlike the 0° case, H-type and relatively metastable domains are formed, resulting in hexagonal to Kagome-like pattern evolutions as a function of the twist angle.[33] Figures 2(a) and (b) show the low-frequency Raman spectra of twisted bilayer $WS_2$ with corrugated structures. The S mode remains only with a twist angle below 3° (above 57°), which corresponds to the moiré domain size of ~6 nm in the rigid limit. Remarkably, the observed S mode shows distinct behavior between 0° and 60° in the relaxed regime. Compared to high-symmetry stackings, the H-type domains have similar peak positions of the S mode. In contrast, the R-type domain has a slightly lower S mode position (~1.0 cm$^{-1}$ with a twist angle of 0.5°). Since the dipole orientations are opposite in inversion domains, the soliton boundaries are movable by an external field.[12] Such sliding of layers affects the S mode position, which might be related to our observation.[32] Further studies are needed to elucidate the exact origin of this behavior. The S mode intensity is stronger than that of the 60° due to the shorter interlayer distance.[20] The S mode position and intensity can be used as a fingerprint of two different reconstruction regimes as demonstrated in Figure 2(g,h). In Figure 2(g), the brighter area (marked as white arrows) corresponds to the angles close to 60° and the darker area (marked as yellow arrows) corresponds to the angles close to 0°.

In the transition regime 1, the shear mode disappears, and a new breathing mode (LB$_2$) emerges. In this regime, the LB$_1$ frequency is blue-shifted, while the LB$_2$ mode appears as the shear mode vanishes. With a shorter moiré period, the area of the high symmetry domain decreases.[29] As a result, the interlayer distance increases, causing the S mode intensity to decrease exponentially. The origin of the new breathing mode is due to the phonon-phonon coupling between the breathing mode of twisted bilayer WS$_2$ and the transverse acoustic phonon of monolayer WS$_2$, which shows an avoided-crossing-like behavior as shown in Figure 2(d).[20,34] As the moiré period decreases, only one LB mode appears in the low-frequency regime, evolving from the LB$_2$ mode of the smaller twisted angles. We categorized this regime as transition regime 2. The breathing mode frequency has a maximum value at 30° and decreases as the twist angle approaches 0° or 60°, owing to the lattice corrugations of the twisted bilayer WS$_2$. The competition between in-plane elastic and interlayer interaction energy leads to the mixing of in-plane and out-of-plane phonons, which is similar to the reported results in other twisted bilayer systems.[20,22,23,29] Figures 2(j) and (k) show the LB$_1$ and LB$_2$ peak positions, respectively. The LB$_2$ position is blue-shifted from 27.2 to 28.3 cm$^{-1}$ as the twist angle changes from 5.5° to 30°, reflecting ~9% variations in interlayer coupling due to corrugation (see Supporting Information, Figure S2). In transition regime 2, the LB$_2$ position represents the twist angles shown in Figure 2(k). We performed similar measurements for stacked MoS$_2$ films (Supporting Information, Figures S3 and S4), which demonstrate the universality of our methods.

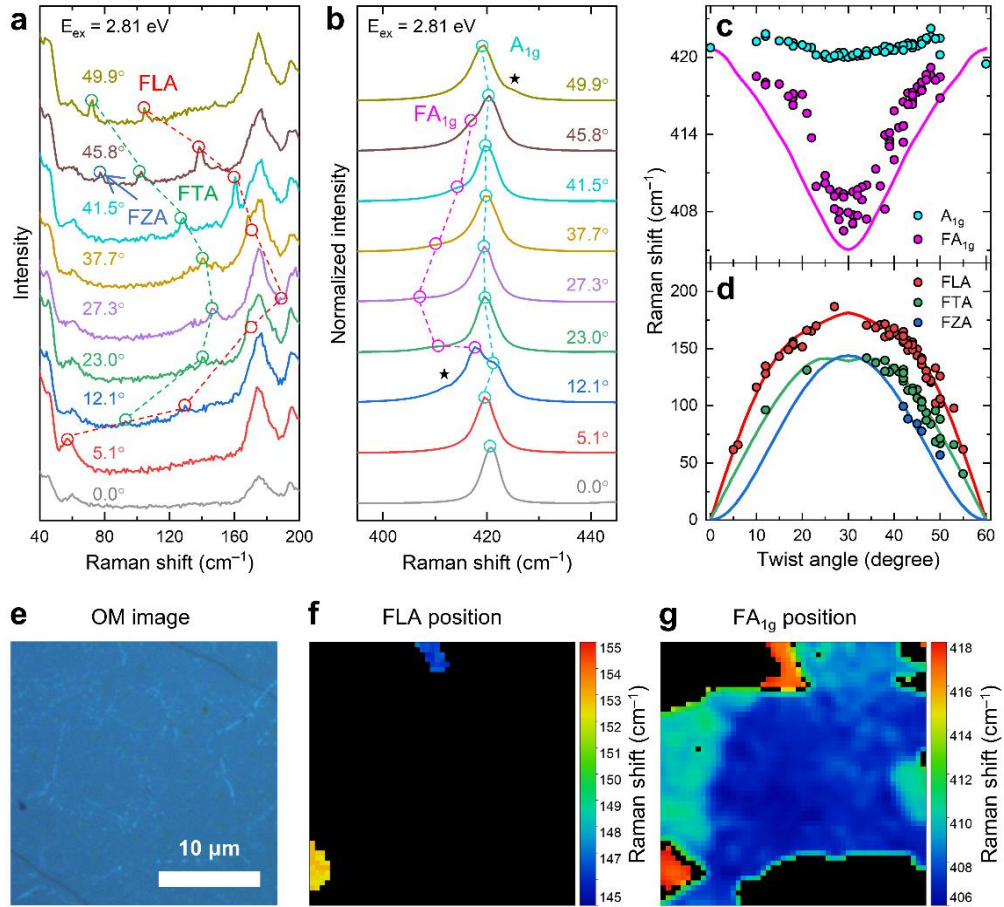

**Figure 3.** Moiré (a) acoustic and (b) optical phonon modes with several twist angles. The typical error for the estimated twist angle is about 0.8°. The Raman peak position (solid circle) compared to the calculated zone-folded phonons (solid lines) from monolayers of $WS_2$ for (c) optical and (d) acoustic phonon branches. (e) OM and Raman image of moiré (f) acoustic and (g) optical phonons.

From the transition regime 2 to the rigid lattice, zone-folded phonons by moiré superlattices are observed. Since the moiré period is larger than the lattice constant of $WS_2$, the Brillouin zone of twisted bilayer $WS_2$ is reduced compared to monolayer $WS_2$, leading to the zone-folding effect in the phonon structure. Phonons with finite momentum in the monolayer will become observable, and such phonons that appear due to the zone-folding effect of the moiré period are called moiré phonons. The direction of the moiré reciprocal vectors depends on the twist angles. To calculate the folded phonon frequencies, we interpolate the phonon dispersions of monolayer $WS_2$ along the Γ–K and Γ–M directions.[19,34] Figure 3(a) shows the Raman spectra of twisted bilayer $WS_2$ with various twist angles from 40 to 200 cm$^{-1}$. Some newly observed phonons have twist angle dependence. To confirm the origin of such angle-dependent phonon modes, we compare the calculated phonon frequencies with experiments shown in Figure 3(c). There could be three acoustic moiré phonons that folded from longitudinal (FLA), transverse (FTA), and out-of-plane (FZA) acoustic phonons. The acoustic moiré phonons are observed from 5° to 55° even though they are weak. Near 30° of twist angle, moiré acoustic phonons are not recognized in the Raman spectra due to overlap with defect-related and second-order Raman modes.

The folded phonon of the $A_{1g}$ branch from the monolayer is observed in the high-frequency range. The $A_{1g}$ mode (419.5 cm$^{-1}$ for 60°) corresponds to out-of-plane vibrations of the sulfur atoms.[35] Figure 3(b) shows the Raman spectra of the $A_{1g}$ mode for several twisted bilayer samples. In addition to the Raman-active $A_{1g}$ mode, new satellite peaks appear in the spectrum. By comparing the angle-dependent peak position with moiré phonons shown in Figure 3(d), we concluded that the peak corresponds to the moiré $A_{1g}$ mode ($FA_{1g}$). This peak is observable in the rigid regime (12° to 48°), which is used to visualize the twist angle in the stacked samples [Figure 3(g)]. Additionally, we have observed several features that cannot be explained by moiré phonons, marked as ★ in Figure 3(b). For example, the peak at ~424 cm$^{-1}$ in the 49.9° twist angle has a higher frequency than that of the $A_{1g}$ mode, which could not be from the zone-folding effect compared to the phonon dispersion of the monolayer. Similar peaks are frequently observed in the transition regime (Supporting Information, Figure S5), suggesting that atomic reconstruction plays a role. In this regime, not only the out-of-plane corrugations but also the in-plane distortions of the lattice are inevitably introduced.[36–38] It modifies the original phonon structures of the constituent layer, making it deviate from the zone-folding effect. In the $MoS_2/WSe_2$ heterostructure, a whirlpool-type in-plane distortion has been observed preferentially in the softer layer.[38] This in-plane distortion breaks the symmetry of the crystal resulting in several new out-of-plane Raman modes depending on the twisted angle. Our observation in the $WS_2$ homostructure could originate from similar effects.

The acoustic moiré phonons are highly sensitive to the twist angle (~6.1 cm$^{-1}$/°). However, due to their weak intensity, they are only applicable to a limited range of twisted angles, as shown in Figure 3(f). In this regard, folded $A_{1g}$ phonons could be used as an alternative indicator of twist angles. The intensity of these modes is three orders of magnitude stronger than moiré acoustic phonons, allowing the visualization of twist angle distribution in the wide range of twist angles. The moiré $A_{1g}$ phonon is sensitive enough to visualize stacked polycrystalline domains (~0.6 cm$^{-1}$/°). The estimated angle sensitivity as a function of the twist angle can be found in Figure S6 (Supporting Information). We have observed domains with a few tens of microns of the moiré superlattices as shown in Figure 3(g).

The semiconducting TMDs show complicated resonant Raman behavior due to the resonance with excitonic states.[39] Depending on the excitation energy, selective enhancement of several Raman modes is reported.[19,39–42] To explore the resonance effect in the twisted system, we have chosen three excitation energies, 1.96, 2.33, and 2.81 eV (632.8, 532.0, and 441.6 nm), which are resonant with A, B, and C excitons, respectively.[41] We found the C exciton resonance is most suitable for the investigation of the stacked vdW films. The details on the excitation energy dependence of the Raman spectrum can be found in Supporting Information Note S2.

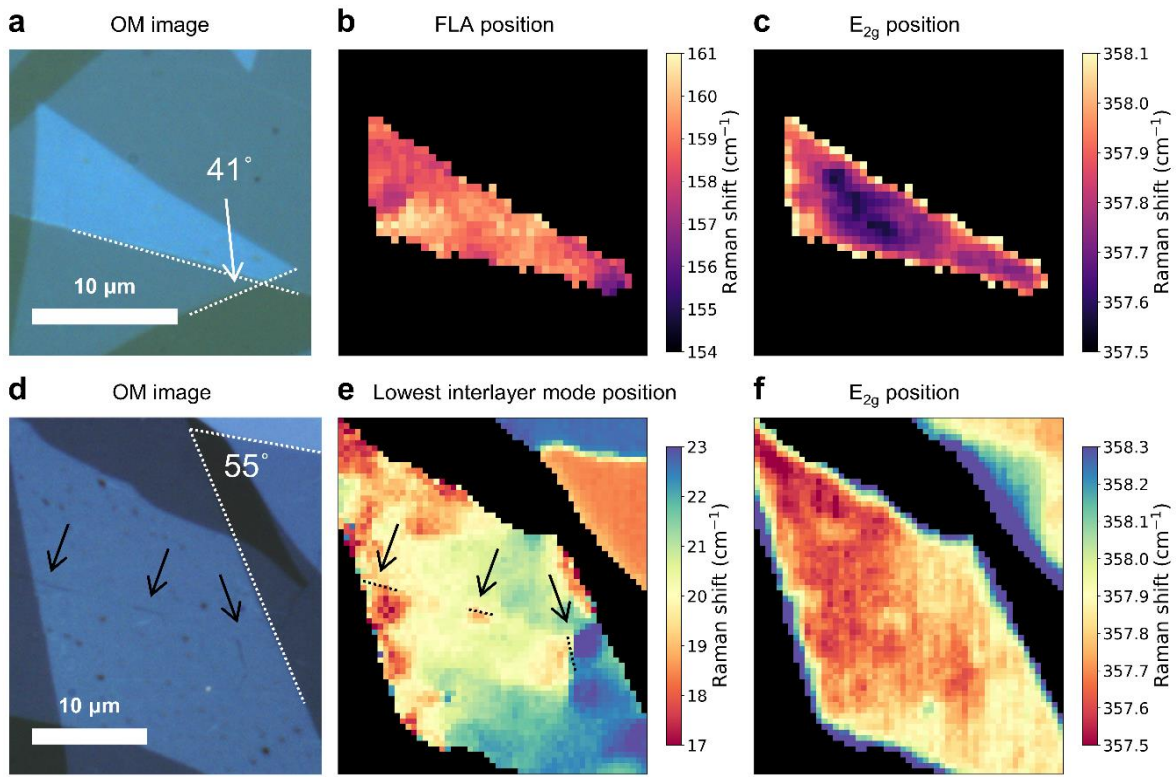

**Figure 4.** The (a) OM, (b) FLA position, and (c) $E_{2g}$ mode position images for the stacked single-crystalline sample in a rigid regime. The (d) OM, (e) lowest interlayer Raman mode

position, and (f) $E_{2g}$ mode position images for the stacked single-crystalline sample in a highly relaxed to transition regime. The cracks are marked as dotted lines.

Not only the stacking configurations but also the in-plane strain distribution can be visualized simultaneously, as shown in Figure 4. The Raman $E_{2g}$ modes (357.8 cm$^{-1}$ for 60°) originate from doubly degenerate vibrations along the in-plane direction, which depend sensitively on the in-plane strain. The Grüneisen parameter for the monolayer WS$_2$ corresponds to 0.54, which resulting in a 1.5 cm$^{-1}$/% peak shift with a Poisson's ratio of 0.22.[43,44] We first investigated the correlation between the stacking structures and mechanical distortions using stacked single-crystalline samples, as shown in Figure 4. Figure 4(a) corresponds to samples in the rigid limit while Figure 4(d) represents the highly relaxed to transition regime. When two monolayer single crystals are stacked, they are expected to have a homogeneous angle distribution. However, the transfer process induces various mechanical distortions in the samples due to thin film mechanics.[10] This occurs because of the thin film/polymer system with a significant mismatch in elastic moduli, where WS$_2$ is much stiffer (~302.4 GPa) than the polymer substrate (2~4 GPa).[44,45] During the pick-up process, the Poisson's effect induces tensile strain which easily results in mud-cracking in the vdW films due to the large moduli mismatch. The positions of the cracks are marked as dotted lines as a guide to the eyes. These cracks further initiate interfacial delamination rather than propagating to the polymer substrate due to weak interfacial adhesion.[46] The delaminated edges seem to be responsible for the angle domains of the fabricated structures, predominantly observed along the cracks or edges of the samples.

Moreover, in the vdW material/polymer system the dominant strain transfer mechanism is shear at the interface. The interfacial shear stress has a spatial distribution along the vdW material when considering the shear-lag effect. It induces an inhomogeneous strain across the vdW material, releasing tensile strain at the edges compared to the inside of the flakes.[47] We observed a similar strain distribution in fabricated stacked vdW films. Figure 4(c) clearly shows the accumulation of tensile strain in the central area of the flakes. We displayed the line profile of the strain distribution across the flake, which can be well-fitted with the shear-lag model (See Supporting Information, Note S3 and Figure S10). This confirms that the residual strains originate from shear-lag effects. The maximum strain observed in this flake is 0.32%, assuming a similar Grüneisen parameter in stacked WS$_2$ as in the monolayer. Since the pick-up process occurs with a moderate tensile strain by Poisson's effect, the upper flakes are transferred with the residual strain, resulting in local strain variations of the stacked samples. These local variations in strain could affect the properties of the fabricated structures and need to be taken into consideration. These effects are significant when the size of the constituent layers is a few tens of microns, exhibiting a gradual variation in twisted angle within ~2° [Figure 5(b)]. The typical scale of the homogeneous angle domains is smaller than ~5 microns, comparable to the elastic stress transfer length in a graphene/polymer system (4~5 microns).[47]

We observed a drastic difference in the strain distribution with different stacking configurations. Compared to the rigid limit, in the highly relaxed and transition regimes, the strain distribution is more correlated with stacking configurations. To visualize this correlation, we obtained the line profile across the flakes as in Figure S11, which clearly demonstrates a correlation between strain and stacking structures. If we consider atomic reconstructions with finite-sized flakes, the strain field will extend to the entire flake starting from the pinning site. Such atomic reconstruction on a mesoscopic scale has been previously reported, resulting in moiré inhomogeneity over few-micron flakes.[15] Thus the observed correlation between stacking configurations and strain is attributed to mesoscopic atomic reconstruction. Additional data from the other samples is shown in Figure S12 (Supporting Information).

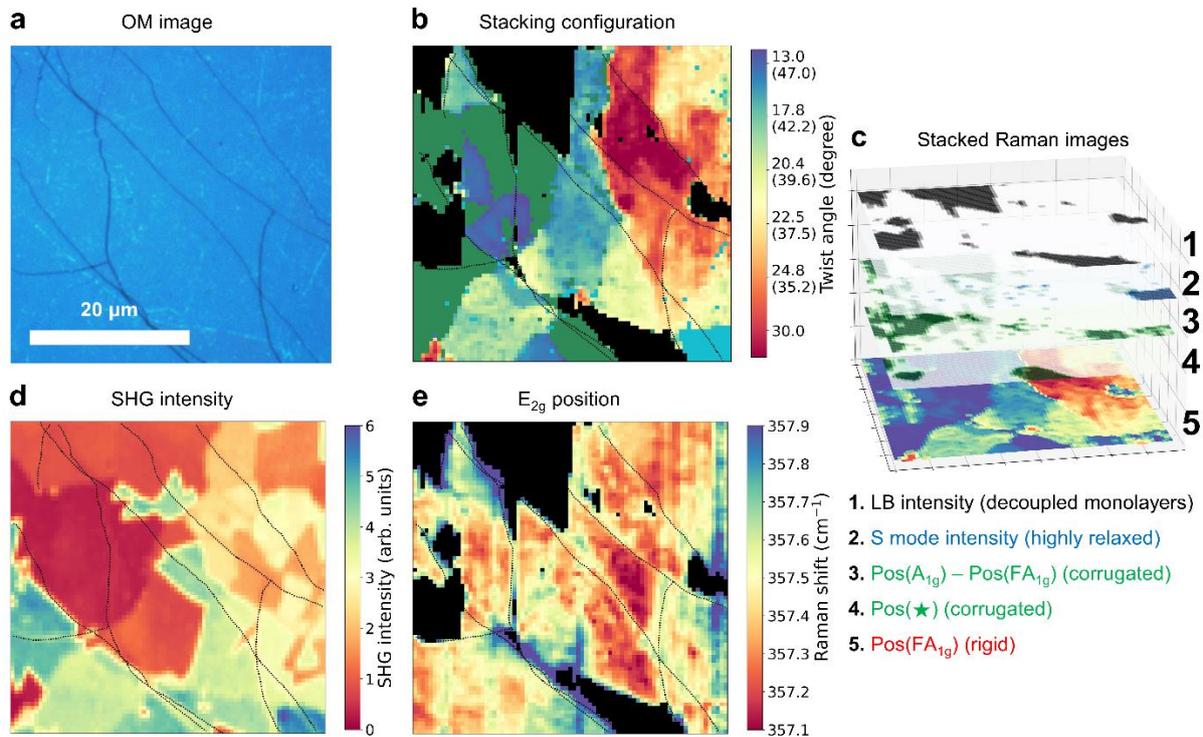

**Figure 5.** The (a) OM image, (b) combined Raman image and its (c) constituent layers. (d) SHG intensity, and (e) $E_{2g}$ mode position images of the stacked polycrystalline film. Five Raman images are combined for (b) as shown in (c). The cracks are marked as dotted lines.

By combining the observed Raman signatures, we have successfully visualized the full stacking configurations of stacked vdW films corresponding to Figure 5(a) as shown in Figure 5(b). We used characteristic signatures of the Raman modes to distinguish different stacking configurations as in Figure 1(c). Figure 5(b) consists of five Raman images to visualize the stacking configurations at all possible twist angles: the LB intensity, the S mode intensity, peak position difference between $A_{1g}$ and $FA_{1g}$, [Pos($A_{1g}$)–Pos($FA_{1g}$)], peak position of the additional $A_{1g}$ mode in Figure 3 [Pos(★)], and peak position of $FA_{1g}$ [Pos($FA_{1g}$)]. Figure 5(c) shows each constituent layer used for Figure 5(b). (also see Supporting Information, Figure S13). Layer 1 corresponds to the LB spectral intensity from 10 to 50 cm$^{-1}$, which indicates the presence of the interlayer interactions. The black area corresponds to decoupled monolayers that lack sufficient interlayer interactions. Layer 2 displays the S mode intensity, indicating highly relaxed regimes. The cyan area corresponds to reconstructed lattices with twist angles below 3° (above 57°), which are similar to R- (H-) type structures. Note that R- and H-type stacking can be distinguished by the S mode intensity as demonstrated in Figure 2. In the present sample, only R-type domains exist. Layers 3 and 4 are used to distinguish the area that is not in the rigid regime; transition 1, transition 2, and the highly relaxed regime. We fitted the $A_{1g}$ mode with two Lorentzian functions for the analysis. In the rigid limit, the peak $A_{1g}$ and $FA_{1g}$ are clearly resolved, otherwise, only a single component of $A_{1g}$ mode will be observed. To identify the presence of the $FA_{1g}$ mode, we calculated the peak position difference from the fitted results. As shown in Figure S14(a), the rigid limit can be identified with a cutoff of 3 cm$^{-1}$, which we have used as layer 3. As mentioned in Figure 3, an additional peak (marked as ★) with higher energy than the $A_{1g}$ mode sometimes appears, which cannot be explained by moiré phonons. This peak appears in the transition regime, so we used Pos(★) as layer 4. Note that the transition 1 regime can be identified by the peak position of $LB_1$ mode as demonstrated in Figure 2(j). The Pos($LB_1$) is not included in Figure 5, as we couldn`t find such an area in the present sample. This layer can be added if one needs to identify the transition 1 regime. The remaining area besides the highly relaxed and transition 1 regime corresponds to the transition 2 regime. Note that the transition 2 regime can also be distinguished by using the peak position of the LB$_2$ mode, as in Figure 2(k). In Figure 5(b), the green area represents the transition 2 regime. The remaining area corresponds to the rigid limit, where the twisted angles are color-coded using moiré A$_{1g}$ phonons. Since the position of the moiré phonon is angle-dependent, Pos(FA$_{1g}$) can be converted into a twist angle. We empirically fitted the Pos(FA$_{1g}$) using a pseudo-Voight function to convert the twist angle from the peak position. The relation between experimental data, calculated phonon dispersion, and the fitting function is shown in Figure S14(b). This demonstrates the successful visualization of the complex stacking configurations of stacked WS$_2$ films covering most of the possible stacked structures. Additional data for other samples can be found in the Supporting Information (Figure S15).

However, in the rigid regime, the zone folding effect dominates, making it impossible to distinguish the mirror pairs (30° + $\theta$ and 30° – $\theta$) by Raman spectroscopy alone. We used SHG spectroscopy as a supporting tool, as it is a sensitive probe for inversion symmetry. For samples close to 0°, the inversion symmetry is broken, resulting in a strong SHG signal. In contrast, for samples close to 60°, the inversion symmetry is preserved, leading to a weak SHG signal (see Supporting Information, Figure S16).[48] The SHG intensity image is shown in Figure 5(d). A similar domain structure has been observed with SHG imaging. At 30°, the SHG intensity is two times stronger than that of the monolayer. The area with SHG intensity stronger (weaker) than 3 corresponds to twist angles from 0° to 30° and from 30° to 60°, respectively. To visualize such a difference, we compared the line profile of SHG intensity and Pos(A$_{1g}$) in Figure S17(d). In domains 1 and 2, the peak position difference is only ~0.7 cm$^{-1}$, which is expected to have ~1.2° rotated domains. However, the SHG intensity changes drastically, as a result, the angle difference in domains 1 (17.4°) and 2 (46.2°) is 28.8°. This shows the important role of SHG imaging as a supportive tool. SHG is also an applicable technique for identifying the crystal structure of moiré materials, but it has limitations when investigating the details of the stacking structure. Especially with reconstruction, the information obtained from SHG measurements will be limited. In this regard, Raman spectroscopy and SHG complement each other and compromise their limitations.

We presented the strain map of stacked polycrystalline films in Figure 5(e). As demonstrated in Figure 4, the shear-lag effect and lattice reconstructions are responsible for the dominant causes of the strains. In polycrystalline stacks, these effects are combined and exhibit complex behaviors. The grain boundaries are expected to have different mechanical properties compared to inner grains, which might affect the local strain distributions in stacked samples.[49] However, we found that the grain boundaries do not significantly affect to the residual strains, the cracks are more responsible. To visualize such effects, we compared the line profiles of SHG intensity and Pos($E_{2g}$) in Figure S17(e). The SHG intensity clearly shows the presence of grain boundaries, while Pos($E_{2g}$) shows negligible changes. The Pos($E_{2g}$) blueshifts near the cracks, suggesting the strain relaxation at the edges due to the shear-lag effect. The estimated value of accumulated strain is about 0.13%. This suggests that even with the presence of grain boundaries, polycrystalline films can be effectively treated as a single sheet mechanically. If one can fabricate the stacked films without cracks and disorders on a large scale, relatively less strain variations are expected inside, which should be beneficial for large-area homogeneous stacked structures. Further studies are needed to prove this idea. The cracks, especially those adjacent to the non-interacting area, exhibit a strong blue shift of the $E_{2g}$ mode, as shown in Figure S17(f). This can be explained by compressive strain induced by the delamination of the cracks and buckling of the thin films due to the presence of disorders.[46] This demonstrates that our technique provides an interesting platform to study and visualize thin film mechanics with atom-thick limits, along with the role of atomic interactions to the stacked structures.

## 3. Conclusion

We investigated stacking reconfigurations on the mesoscopic scale when polycrystalline vdW films are stacked. Thanks to the versatility of Raman spectroscopy, we were able to visualize complicated stacking configurations, enabling us to study correlations with mechanical distortions in the stacked samples. The transfer process contributes to residual strain which can be understood through thin film mechanics with a large elastic property mismatch. With atomic corrugations, we found that the in-plane strain fields are more closely correlated with the stacking configuration, highlighting the significance of atomic reconstruction on the mesoscopic scale. Furthermore, we found that the role of grain boundaries is not crucial in polycrystalline films, instead cracks are responsible for the inhomogeneous strain. Our work provides an invaluable characterization tool for studying moiré materials and provides a better understanding of mechanical distortions in the stacked vdW system.

## 4. Methods

*Sample preparation of monolayer TMD samples*

The monolayer $WS_2$ samples are prepared by either mechanical exfoliation or MOCVD. The monolayer WS$_2$ is exfoliated from a bulk crystal (HQ Graphene) on commercially available PDMS (GelPak) using Scotch tape. The monolayer WS$_2$ on PDMS is then transferred to an empty fused silica substrate by stamping PDMS. WS$_2$ (and MoS$_2$) films were synthesized using a home-built MOCVD with a 2-inch quartz tube furnace. Tungsten (molybdenum) hexacarbonyl (Sigma Aldrich) and diethyl sulfide (Sigma Aldrich) were used as precursors and mixed with a constant pressure of Ar with 800 Torr. The flow of precursors was controlled by mass flow controllers (MFCs, Alicats). The flow of Ar and H$_2$ gases is controlled individually by MFCs. The growth temperature was kept at 700°C, and NaCl was placed upstream of the furnace to promote growth. WS$_2$ was directly grown on the Si wafer with 300 nm of SiO$_2$, and the typical growth area was a few centimeters.

*Modified dry stamping method*

To fabricate the stacked vdW films, we used a modified dry stamping method by PDMS. We injected one drop of the ethanol while the PDMS are in contact with vdW materials to increase the transfer yield.[24,25] The detached vdW films on the PDMS were transferred onto an empty fused silica substrate and the process was repeated to transfer another layer on top.

*Raman spectroscopy*

The Raman measurements were performed using a home-built microscope. The excitation sources were the 441.6, 532.0, and 632.8 nm (2.81, 2.33, and 1.96 eV) lines of the He-Cd laser (Kimmon), diode pumped solid state laser (Laser Quantum, Torus), and He-Ne laser (Pacific Lasertec), respectively. The laser beam was focused onto the sample by a 100× microscope objective lens (0.90 N.A.), and the scattered light was collimated by the same objective. The laser power was kept below 100 μW to prevent damage to the samples. The Schmidt-Czerny-Turner (SCT) spectrometer (Princeton Instruments, IsoPlane320) with a 2400 grooves/mm grating was used and the signal is detected with either an sCMOS (Princeton Instruments, Kuro2048B) or a back-illuminated charge-coupled device (Princeton Instruments, PIXIS 400BRX) detector. Bragg notch filters (Optigrate) were used to access the low-frequency range down to 5 cm$^{-1}$. For Raman imaging, the samples were raster scanned with exposure times of less than 10 seconds at each point.

*SHG measurements*

The ultrafast mode-locked fiber laser (Femtopro IR, Toptica) or the picosecond supercontinuum laser (NKT photonics, SuperK Extreme) coupled with a wavelength selector

(Photon etc., LLTF Contrast) are used as excitation sources. We chose the non-resonant excitation wavelength (1140 or 1560 nm) for the SHG measurements to avoid local variation within a single domain. The samples are raster-scanned for SHG imaging.

**Supporting Information**

Supporting Information is available from the Wiley Online Library or from the author.

**Acknowledgements**

This work was supported by the National Research Foundation (NRF) grants funded by the Korean government (MSIT) (NRF-2020R1C1C1005963, NRF-2021R1A4A1032085, NRF-2022M3H3A1063074).

Received: ((will be filled in by the editorial staff))
Revised: ((will be filled in by the editorial staff))
Published online: ((will be filled in by the editorial staff))

*Yoon Seong Heo, Tae Wan Kim, Wooseok Lee, Jungseok Choi, Soyeon Park, Dong-Il Yeom, and Jae-Ung Lee\**

**Mesoscopic Stacking Reconfigurations in Stacked van der Waals Film**

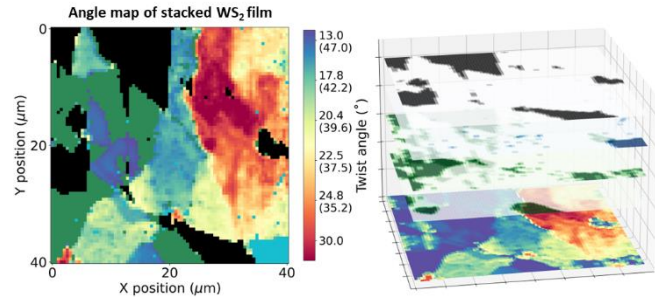

Mesoscopic-scale stacking reconfigurations and their correlations with mechanical distortions are investigated when van der Waals films are stacked. The origins of deformations are revealed from either the transfer process or the extended impact of atomic reconstructions.

# Supporting Information

**Mesoscopic Stacking Reconfigurations in Stacked van der Waals Film**

*Yoon Seong Heo, Tae Wan Kim, Wooseok Lee, Jungseok Choi, Soyeon Park, Dong-Il Yeom, and Jae-Ung Lee\**

**Table S1.** Summary of Raman studies of twisted TMD bilayers.

| Twisted bilayer | $\lambda_{ex}$ | Characteristic Raman modes | | | | | | |
|---|---|---|---|---|---|---|---|---|
| | | Reconstructed shear mode | | Breathing mode $LB_1$ | Breathing mode $LB_2$ | Moiré acoustic | Moiré optical | Main modes ($E_{2g}$ and $A_{1g}$) |
| | | R-type | H-type | | | | | |
| $MoS_2$[1] | 2.33 eV | | | | | | | ○ |
| $MoS_2$[2] | 2.33 eV | | | | | | ○ | ○ |
| $MoS_2$[3] | 2.33 eV | ○ | | ○ | ○ | | | ○ |
| $MoS_2$[4] | 1.83 to 2.81 eV | | | | | ○ | ○ | ○ |
| $MoSe_2$[5] | 2.33 eV | | ○ | ○ | ○ | | | |
| $MoSe_2$[6] | 2.33 eV | | | | | ○ | | |
| $WSe_2$[7] | 2.33 eV | ○ | ○ | | | ○ | ○ | |
| $WS_2$[8] | 2.54 eV | | | | | | | ○ |
| $WS_2$ (This work) | 1.96, 2.33, and 2.81 eV | ○ | ○ | ○ | ○ | ○ | ○ | ○ |

**Note S1. Identification of relative angles in stacked single crystalline $WS_2$.**

Two single crystalline flakes with a domain size of ~10 microns were stacked, and polarized SHG measurements were performed to determine the twisted angles. The analyzer was fixed parallel to the incident polarization, and the incident polarization angles were rotated with respect to the crystal orientation using a half-waveplate. The polarized SHG exhibited the six-fold behavior in which the maximum intensity aligned with the armchair direction. (Figure S1).[9] The triangular edges from the OM images were used to identify the mirror pair (0° to 30° and 60° to 30°), which could not be distinguished by polarized SHG due to six-fold symmetry. The typical error bar for angle estimation was 0.8°, obtained by comparing the OM and polarized SHG measurements. This error is consistent with the values reported in other literatures.

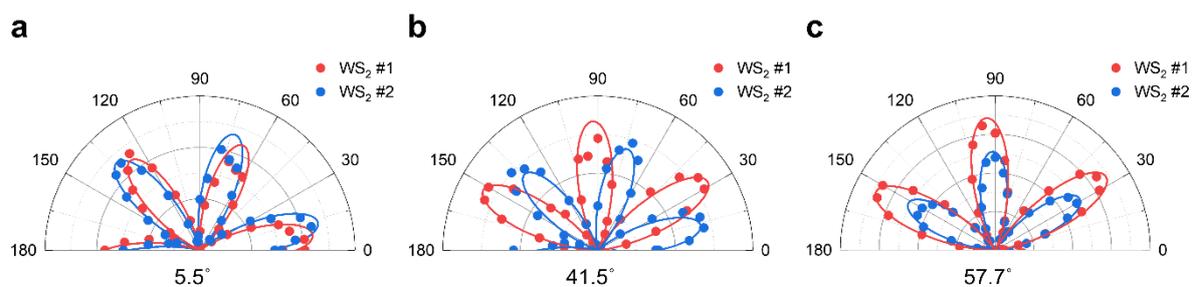

**Figure S1.** Three representative results of the polarized SHG measurements on stacked single crystalline WS$_2$. To distinguish the twist angle 30° + $\theta$ and 30° − $\theta$, the triangular edges of the OM images are used. The twist angle is determined as (a) 5.5°, (b) 41.5°, and (c) 57.7°.

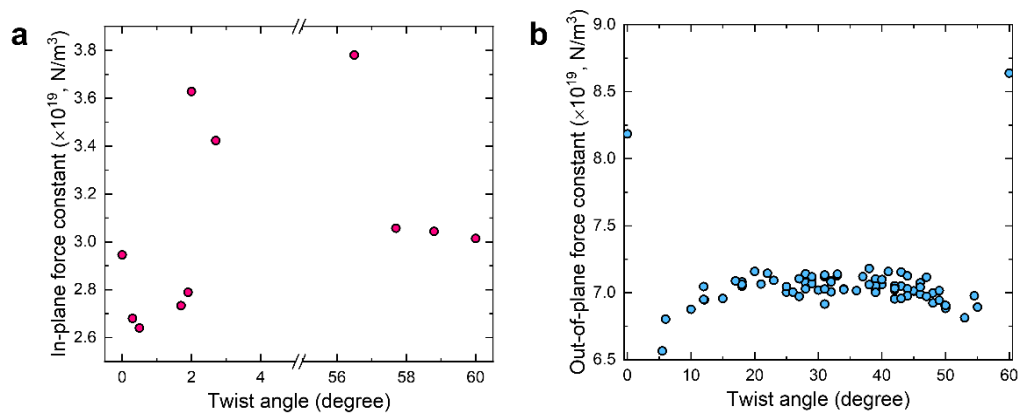

**Figure S2.** The (a) in-plane and (b) out-of-plane interlayer force constants as a function of the twist angle calculated by the linear chain model.[10,11]

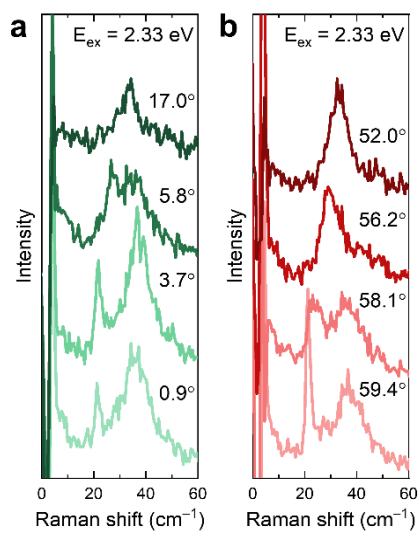

**Figure S3.** Low-frequency Raman spectra of twisted bilayer MoS$_2$ under 2.33 eV excitation. Twist angles from (a) 0.9° to 17.0° and (b) 52.0° to 59.4°. The twist angle is measured by using optical images of samples and has a typical error of ~2°. The evolution of interlayer modes is similar to that of twisted bilayer WS$_2$.

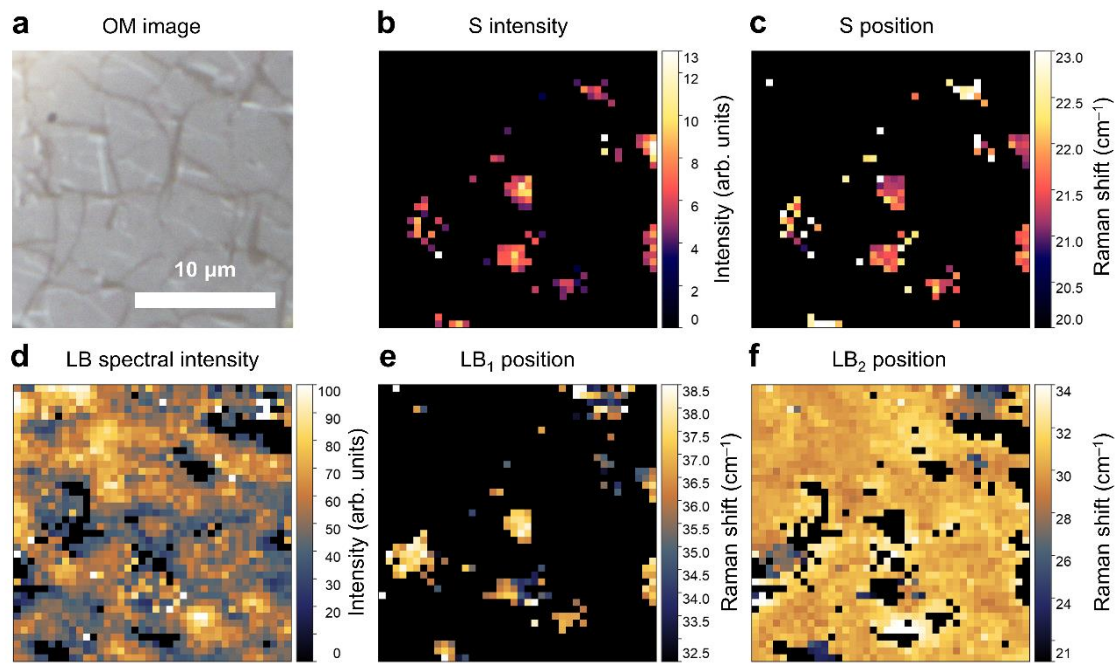

**Figure S4.** (a) The optical image of stacked MoS$_2$ film. The Raman images of (b) S mode intensity, (c) S mode position, (d) LB spectral intensity, (e) LB$_1$ position, and (f) LB$_2$ position, respectively.

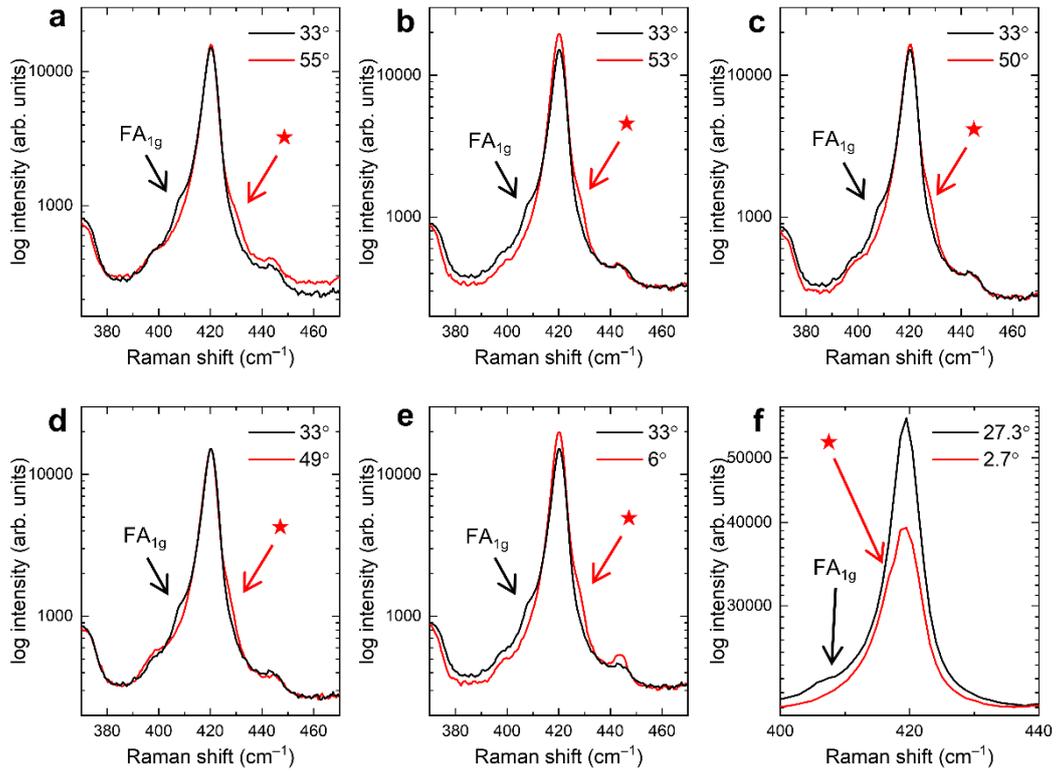

**Figure S5.** Raman spectra of samples with peak ★ (red solid line). The twist angle corresponds to (a) 55°, (b) 53°, (c) 50°, (d) 49°, (e) 6°, and (f) 2.7°, respectively. For comparison, Raman spectra without ★ peak are shown by a black solid line. The peak ★ is usually observed in the transition regime (below 12° and above 48°) with 2.81 eV excitation.

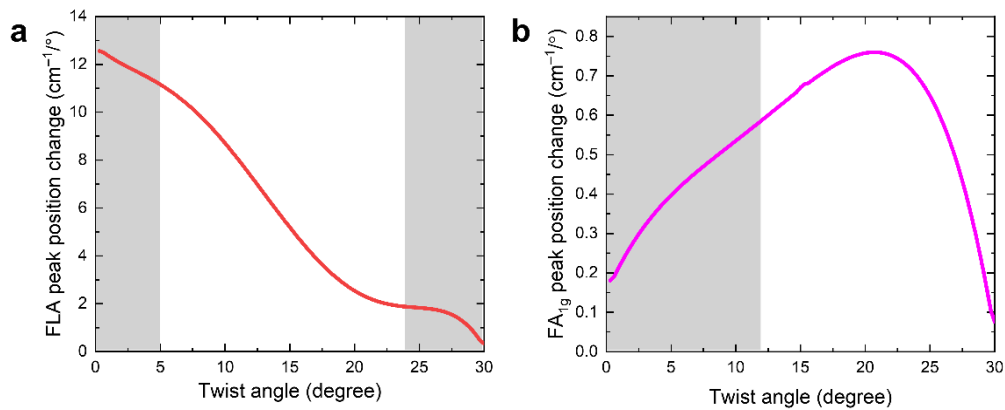

**Figure S6.** The peak position change per degree of (a) FLA and (b) $FA_{1g}$ phonon. These are calculated by differentiating the corresponding moiré phonon dispersion in Figure 3(c) and (d). The gray-shaded area represents the angle range that the moiré phonon is not observed.

**Note S2. Resonance effect on the Raman spectra of stacked vdW materials.**

We have measured the Raman spectra with different excitation energies. The semiconducting TMDs show complicated resonant Raman behavior due to the interaction with excitonic structures.[12] With excitation energy, selective enhancement of several Raman modes is reported.[4,11–14] We have chosen three excitation energies, 1.96, 2.33, and 2.81 eV (corresponding to 632.8, 532.0, and 441.6 nm), which are resonant with A, B, and C excitons, respectively.[11] The Raman spectra in Figures 1, 2, and 3 are measured with 2.81 eV, and Figure S7 shows the spectra with 1.96 and 2.33 eV.

With A exciton resonance, a strong luminescence background appears and hinders many of the Raman signatures. Some features appear, but a systematic dependence on the twisted angle could not be found in the low-frequency regions [Figure S7(a–c)]. Furthermore, the acoustic moiré phonons are hard to observe at this excitation energy. Only some defect-related Raman modes are visible at 147.5 cm$^{-1}$ [Figure S8(a)].[15] The splitting of the $A_{1g}$ mode is observed for the A exciton resonance, however, it shows no twist angle dependence [Figure S8(b)]. This can be explained by the infrared-active $A_{2u}$ mode of $WS_2$.[16] The separation between $A_{1g}$ and $A_{2u}$ peaks is ~3.2 cm$^{-1}$ similar to the values from H- and R- type stackings. For this vibration mode, the bottom sulfur atom at the upper layer and the top sulfur atom at the lower layer move in the same direction. Considering the vibrational patterns of the $A_{2u}$ mode, this peak should be insensitive to the interlayer interactions, which rationalizes our observation. The information about the $A_{2u}$ peak including its peak position is shown in Figure S9.

The luminescence background is weaker than the A exciton resonance with the B exciton resonance. A new peak at ~24.4 cm$^{-1}$ was observed, which cannot be explained by interlayer phonon modes [Figure S7(d–f)]. These modes are related to acoustic phonon branches.[17] With different twist angles, there are some variations in the spectral shapes with no systematic changes. Comparing this variation with theoretical calculations may lead to the origin of these peaks, suggesting that the observed features are not closely related to interlayer interactions. Localized intra-layer excitons confined in the monolayer could play a role. In the case of B exciton resonance, some of the moiré acoustic phonons are visible although it is weak [Figure S8(c)]. Similar features with A exciton resonance that can be related to the defect are found in the Raman spectra. The $A_{2u}$ mode is also observed with the B exciton resonance [Figure S8(d)]. The moiré optical phonons are hard to recognize with B exciton resonance. The moiré phonons and interlayer Raman modes are clearly observed with C exciton resonance similar to $MoS_2$.[4] This makes the excitation energy close to the C exciton energy (~2.8 eV) suitable to visualize the stacking configurations in the TMDs moiré structures.

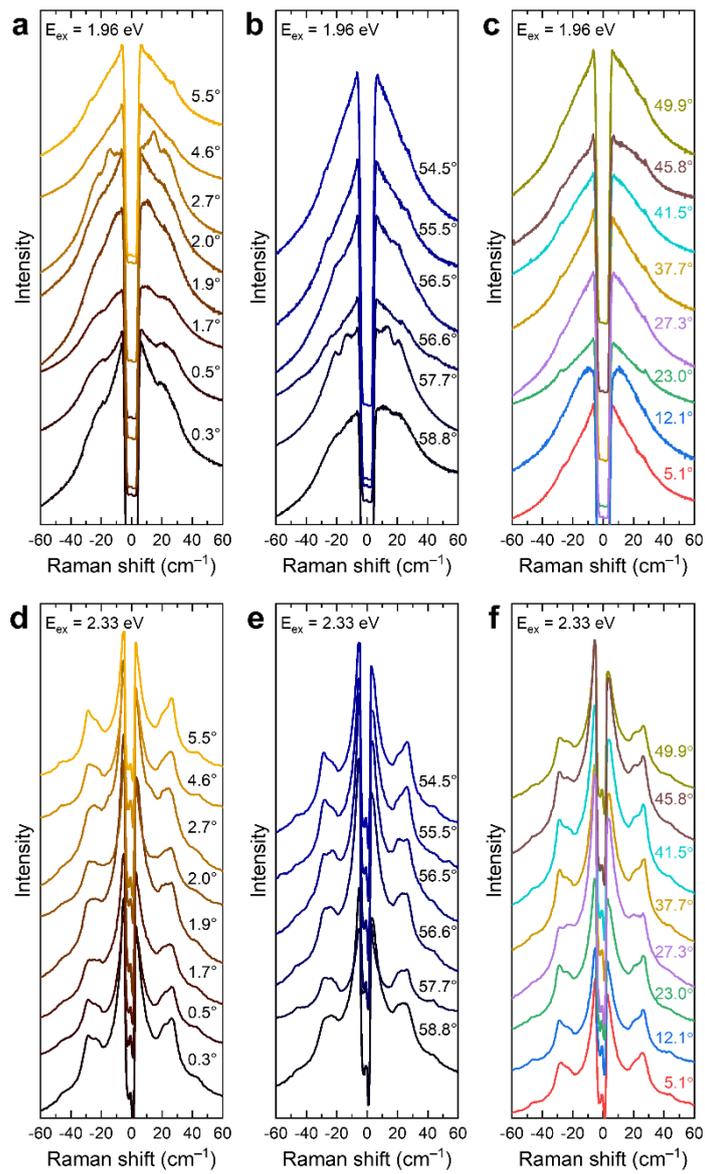

**Figure S7.** Low-frequency Raman spectra of twisted bilayer WS$_2$ under 1.96 and 2.33 eV excitations. Twist angles from (a) 0.3° to 5.5°, (b) 54.5° to 58.8°, and (c) 5.1° to 49.9° with 1.96 eV excitation and (d–f) 2.33 eV excitation. The typical error for the estimated twist angle is about 0.8°.

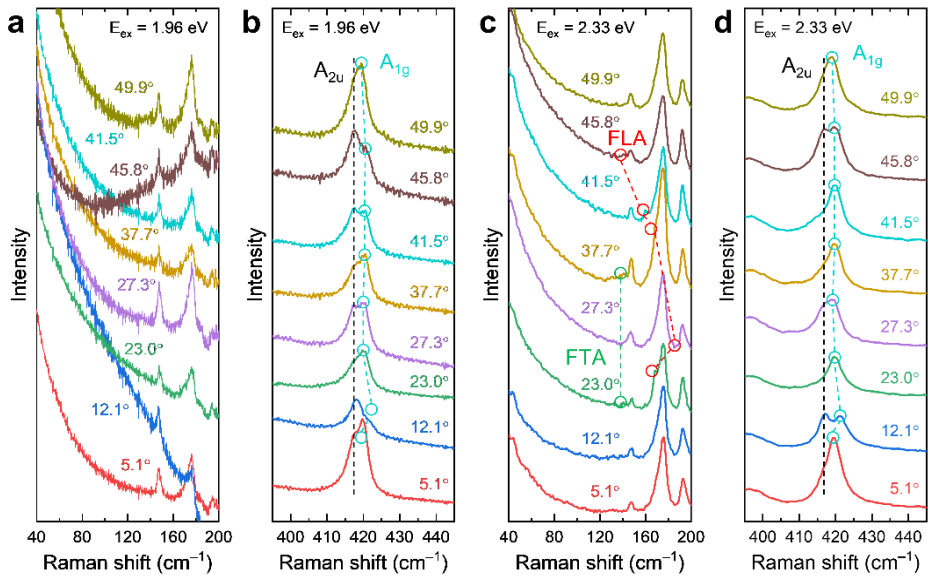

**Figure S8.** The moiré (a,c) acoustic and (b,d) optical phonons with (a,b) 1.96 and (c,d) 2.33 eV excitation energies. Twist angles from 5.1° to 49.9°. The typical error for the estimated twist angle is about 0.8°.

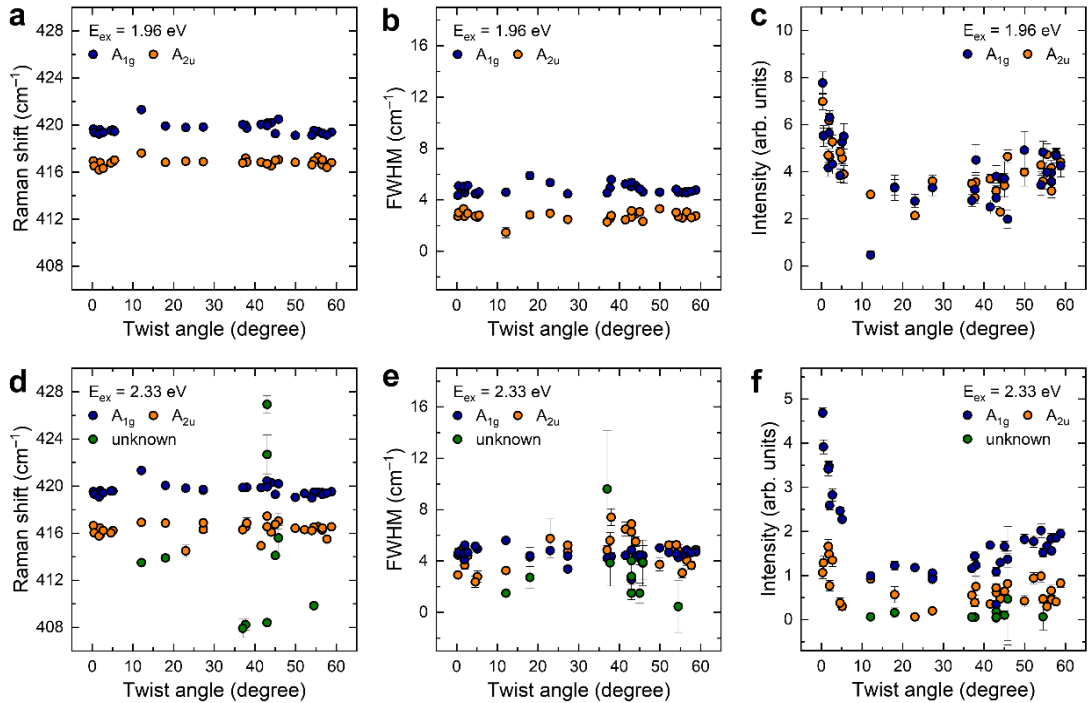

**Figure S9.** Raman-active $A_{1g}$ mode and infrared-active $A_{2u}$ mode of twisted bilayer $WS_2$. (a,d) The peak position, (b,e) full-width half maximum (FWHM), and (c,f) intensity with 1.96 eV and 2.33 eV excitations, respectively. The peaks assigned as unknown in (d–f) are not observed with 1.96 and 2.81 eV excitations.

**Note S3. Shear-lag model analysis**

We used the shear-lag model to analyze the strain distribution in stacked single-crystalline $WS_2$ films with rigid limit. The transfer process for vdW films using polymer substances, such as PDMS, inevitably introduces unwanted strain on the samples. During the press and pilling off processes, shear stress is applied between the vdW film and PDMS interface due to the shear-lag effect. The transferred film exhibits strain distribution that is inhomogeneous on a micron scale.

In the situation where the interfacial shear stress is transferred across the interface from PDMS to the vdW film, the strain distribution $\varepsilon$ in vdW film will be given as followings:[18]

$$\varepsilon(x) = \varepsilon_0 \left[ 1 - \frac{\cosh \beta x}{\cosh \frac{\beta L}{2}} \right] \quad (S1)$$

Here, the $\varepsilon_0$ is the applied strain to the PDMS, $\beta$ is the stress transfer efficiency, and $L$ is the half of length of the vdW film along the $x$ direction. According to this equation, strain is maximally accumulated at the center of the sample. We compared this model with our experimental data from Figure 4(c). The results are shown in Figure S10, which explains the origin of inhomogeneous strain distribution in our sample. It is worth noting that the strain distribution in samples in a highly relaxed regime cannot be explained by this model only (Figure S11). This demonstrates that the dominant mechanism for strain distribution is contrasting in these two structural limits.

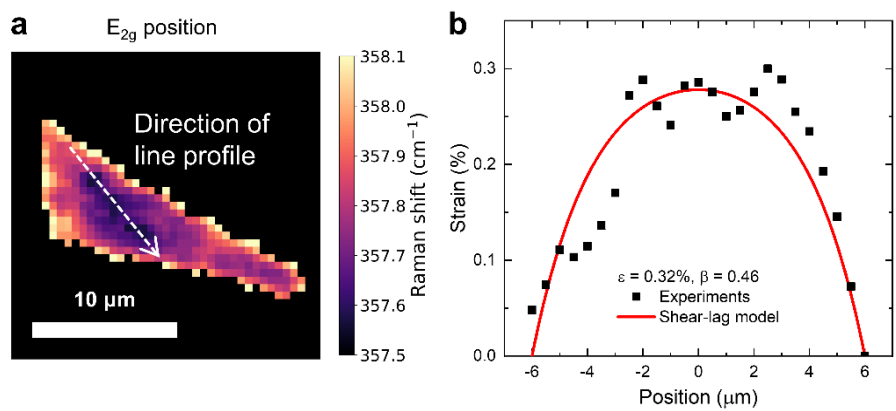

**Figure S10.** (a) The $E_{2g}$ position image for the stacked single-crystalline sample in a rigid regime. The white dashed arrow shows the direction of the line profile in (b). (b) Experimentally obtained strain level and comparison with the shear-lag model.

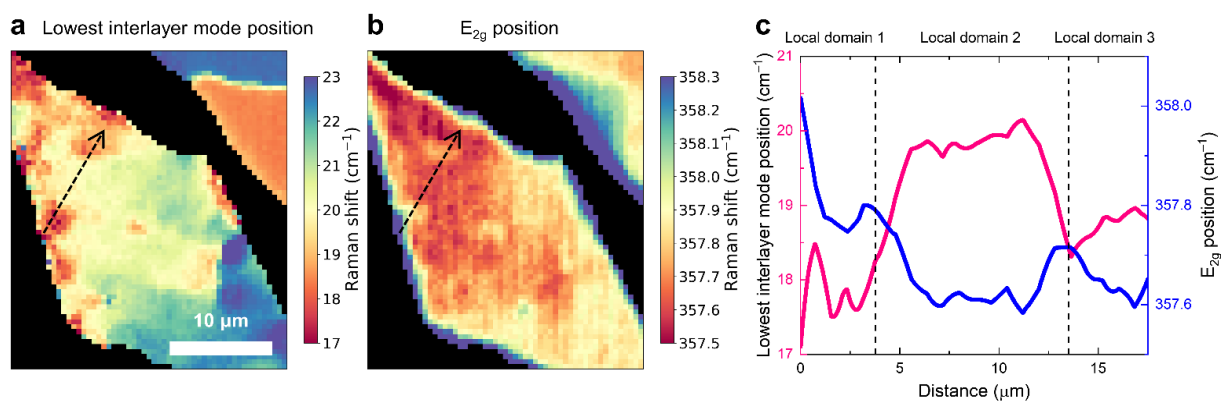

**Figure S11.** The (a) lowest interlayer mode position and (b) $E_{2g}$ position images for the stacked single-crystalline sample in Figure 4(d) which is in a highly relaxed to transition regime. The Black dashed arrow shows the direction of the line profile in (c). (c) Line profiles of lowest interlayer mode position (pink) and $E_{2g}$ position (blue). Black dashed lines indicate the boundaries between locally twisted domains in the stacked single-crystalline sample.

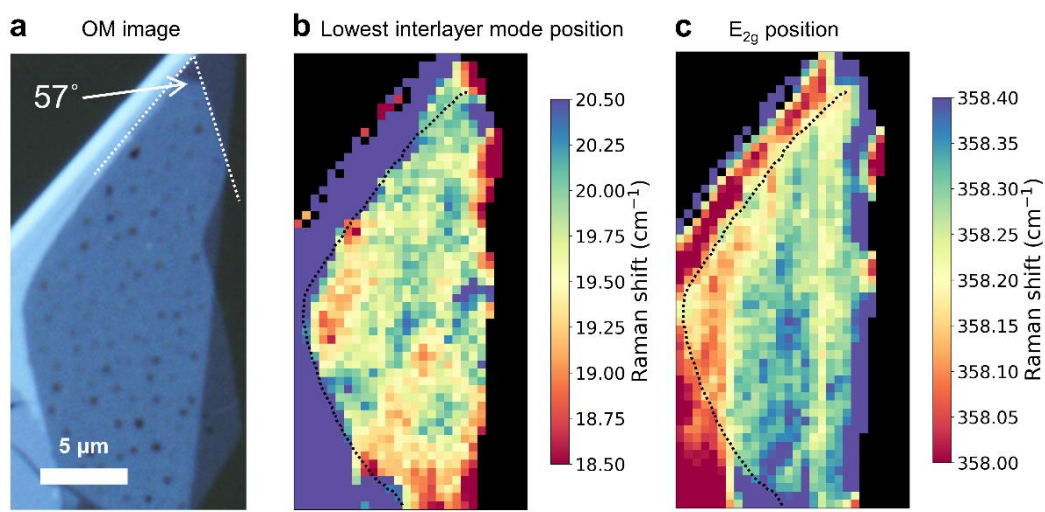

**Figure S12.** The (a) OM, (b) lowest interlayer Raman mode position, and (c) $E_{2g}$ mode position images for the stacked single-crystalline sample in a highly relaxed regime.

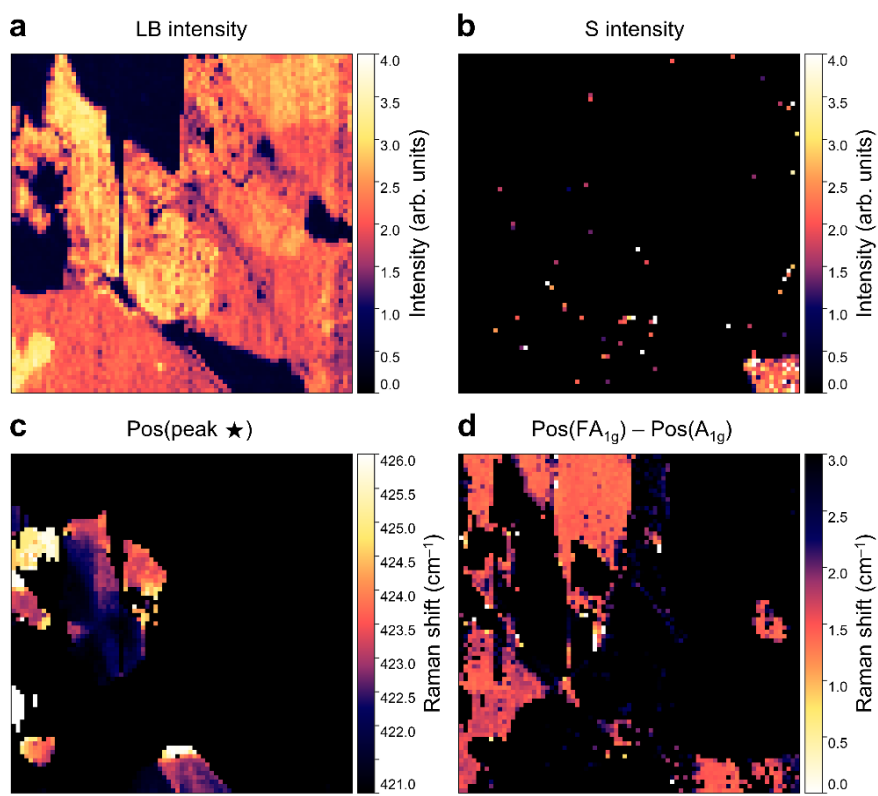

**Figure S13.** Four spectroscopic images are used in Figure 5(b). (a) LB intensity, (b) S intensity, (c) peak position of ★ which is mentioned in the main text and Figure 3(b), and (d) peak position difference between $FA_{1g}$ and $A_{1g}$.

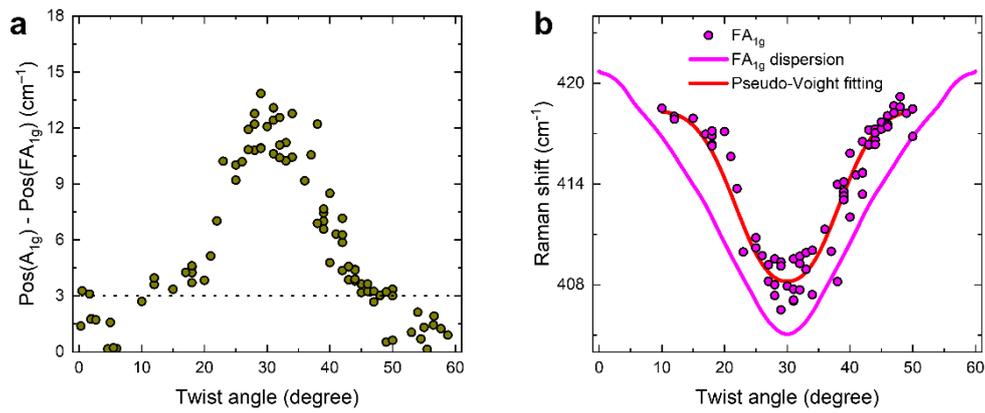

**Figure S14.** (a) The difference between the peak positions of $A_{1g}$ and $FA_{1g}$. The dashed line corresponds to a value of 3 cm$^{-1}$. (b) The experimental peak position of $FA_{1g}$, the calculated $FA_{1g}$ phonon dispersion, and the pseudo-Voight fitting of the experimental data.

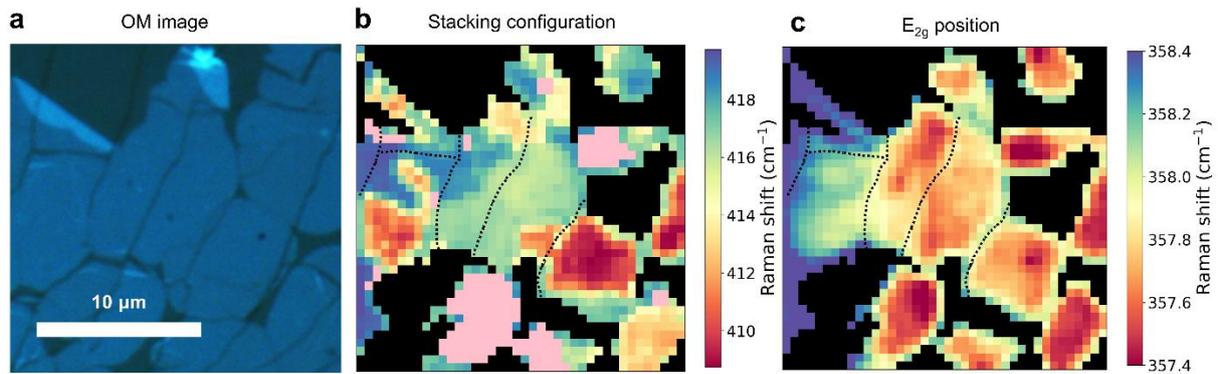

**Figure S15.** The (a) OM, (b) combined Raman, and (c) $E_{2g}$ mode position images for the stacked polycrystalline sample. Three Raman images are combined in (b) [S intensity (pink area), LB intensity (black area for uncoupled or monolayer region), and $FA_{1g}$ position (color-coded)] similar to the image in Figure 5(b).

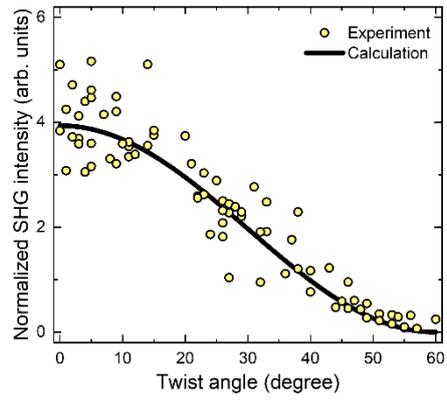

**Figure S16.** Angle-dependent SHG intensity of twisted bilayer WS$_2$. The calculation is done by considering the vector superposition of the second harmonic electric field of each layer.[19]

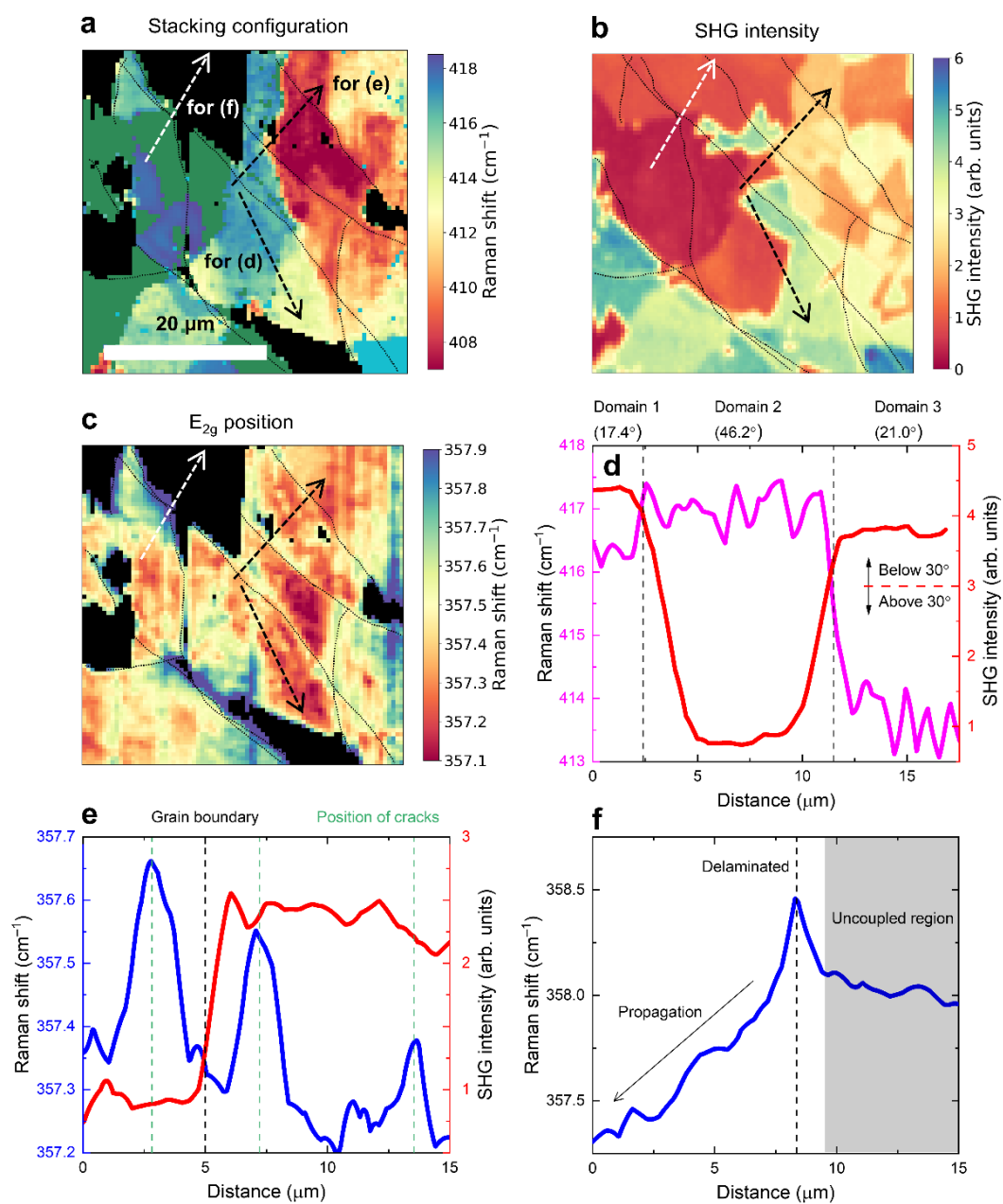

**Figure S17.** The (a) stacking configuration, (b) SHG intensity, and (c) $E_{2g}$ position images of the stacked polycrystalline film in Figure 5(a). Black and white dashed arrows correspond to the direction of line profiles drawn in (d–f). Line profiles of (d) $FA_{1g}$ position (magenta) and SHG intensity (red), (e) $E_{2g}$ position (blue) and SHG intensity (red), and (f) $E_{2g}$ position.